\documentclass[manuscript, screen]{acmart}

\usepackage{booktabs} 

\begin{document}
\title{The SAGE Project: a Storage Centric Approach for Exascale Computing}

\author{Sai Narasimhamurthy}
\affiliation{%
  \institution{Seagate Systems UK}
}

\author{Nikita Danilov}
\affiliation{%
  \institution{Seagate Systems UK}
}

\author{Sining Wu}
\affiliation{%
  \institution{Seagate Systems UK}
 }

\author{Ganesan Umanesan}
\affiliation{%
  \institution{Seagate Systems UK}
}

\author{Steven Wei-der Chien}
\affiliation{%
  \institution{KTH Royal Institute of Technology}
} 

\author{Sergio Rivas-Gomez}
\affiliation{%
 \institution{KTH Royal Institute of Technology}
 }
 
 \author{Ivy Bo Peng}
\affiliation{%
 \institution{KTH Royal Institute of Technology}
 }
 
\author{Erwin Laure}
\affiliation{%
  \institution{KTH Royal Institute of Technology}
}

\author{Shaun de Witt}
\affiliation{%
  \institution{Culham Center for Fusion Energy}
 }

\author{Dirk Pleiter}
\affiliation{\institution{J\"ulich Supercomputing Center}}

\author{Stefano Markidis}
\affiliation{\institution{KTH Royal Institute of Technology}}

\renewcommand{\shortauthors}{S. Narasimhamurthy et al.}

\begin{abstract}
SAGE (Percipient \textbf{S}tor\textbf{AG}e for \textbf{E}xascale Data Centric Computing) is a European Commission funded project towards the era of Exascale computing. Its goal is to design and implement a Big Data/Extreme Computing (BDEC) capable infrastructure with associated software stack. The SAGE system follows a \emph{storage centric} approach as it is capable of storing and processing large data volumes at the Exascale regime.

SAGE addresses the convergence of Big Data Analysis and HPC in an era of next-generation data centric computing. This convergence is driven by the proliferation of massive data sources, such as large, dispersed scientific instruments and sensors where data needs to be processed, analyzed and integrated into simulations to derive scientific and innovative insights. A first prototype of the SAGE system has been been implemented and installed at the J\"ulich Supercomputing Center. The SAGE storage system consists of multiple types of storage device technologies in a multi-tier I/O hierarchy, including flash, disk, and non-volatile memory technologies. The main SAGE software component is the Seagate Mero Object Storage that is accessible via the Clovis API and higher level interfaces. The SAGE project also includes scientific applications for the validation of the SAGE concepts.

The objective of this paper is to present the SAGE project concepts, the prototype of the SAGE platform and discuss the software architecture of the SAGE system.
\end{abstract}

%
%
%

\keywords{SAGE Project, Storage Centric Computing, Exascale Computing}

\maketitle

\section{Introduction}
The next-generation Exascale supercomputers will be available to the community sometime in the 2021-2023 timeframe and it will be capable of delivering up to $10^{18}$ floating point operations per seconds to support traditional HPC compute-intensive applications. With the emergence of new HPC data centric applications, such as workflows and data analytics workloads, the definition of Exascale computing is now broadening to include  storage and processing of an order of an exabyte of data. In fact, Big Data Analysis and HPC are converging as massive data sets, such as very large volumes of data from scientific instruments and sensors, needs to be processed, analyzed and integrated into simulations. For this reason, we envision that Exascale supercomputers will be capable to be exploited by applications and workflows for science and technological innovation. 

Computing infrastructure innovation has been driven by Moore's law and the development of even more parallelism with multi-core, many-core and accelerator processing to accommodate the increasing performance requirements of Exascale. However I/O and storage have lagged far behind in computational power improvement. Storage performance in the same time period is predicted to have improved for only 100 times, according to early estimates provided by Vetter et al.~\cite{vetter2009hpc}. In fact, at the time of publication of this work, the performance of disk drives per unit capacity is actually decreasing with new very high capacity disk drives on the horizon~\cite{vetter2009hpc}. Simultaneously, the landscape for storage is changing with the emergence of new storage device technologies, such as flash (available today) and the promise of non-volatile memory technologies available in the near future~\cite{peng2016exploring, peng2017exploring}. The optimal use of these devices (starting with flash) in the I/O hierarchy, combined with existing disk technology, is only beginning to be explored in HPC~\cite{FastForward} with burst buffers~\cite{liu2012role}. 

The SAGE project is a European Commission funded project to investigate Exascale computing~\cite{narasimhamurthy2018sage}. It supports a storage centric view of Exascale computing by proposing hardwares to support multi-tiered I/O hierarchies and associated software. They provide a demonstrable path towards Exascale. Further, SAGE proposes a radical approach in extreme scale HPC by moving traditional computations, typically done in the compute cluster, to the storage system. This has the potential of significantly reducing the energy footprint of the overall system~\cite{reed2015exascale}. 

The primary objective of this paper is to present the initial hardware and software architecture of the SAGE system. The paper is organized as follows. Section~\ref{sec-challenges} presents the initial development of the SAGE platform. Section~\ref{sec-architecture} describes the SAGE platform architecture and software stack. Section~\ref{sec-relwork} discusses the related work. Finally, Section~\ref{sec-conclusions} summarizes the paper and outlines the future work.

\section{A Storage Centric Architecture}
\label{sec-challenges}
The SAGE storage system developed by the SAGE consortium provides a unique paradigm to store, access and process data in the realm of extreme-scale data centric computing.  

The SAGE platform consists of multiple tiers of storage device technologies. At the bottom of the stack is the \emph{Unified Object-Based Storage Infrastructure}. The system does not require any specific storage device technology type and accommodates  upcoming NVRAM, existing flash and disk tiers.  For the NVRAM tier, we are using Intel 3D XPoint technology~\cite{bourzac2017has} in our \emph{Tier-1}. We will also use emulated NVDIMMs (Non-Volatile DIMMs) in Tier-1 because of the lack of NVDIMM availability in vendor roadmaps. We are using Flash based solid state drives in \emph{Tier-2}. Serial Attached SCSI high performance drives are contained in \emph{Tier-3} and archival grade, high capacity, slow disks ( based on Serial ATA and Shingled Magnetic Recording) are contained in \emph{Tier-4}. These tiers are all housed in standard form factor enclosures that provide their own computing capability through standard x86 embedded processing components, which are connected through an FDR Infiniband network. Moving up the system stack, compute capability increases for faster and lower latency device tiers. The storage system is also capable to perform computation in storage (either through a function shipping interface or a run-time supporting, e.g.,pre-/post-processing of data) on behalf of the applications. This avoids the need to move data back and forth between compute subsystems and storage subsystems as in a typical HPC cluster.

\begin{figure}
  \begin{center} 
  \includegraphics[width=0.9\linewidth]{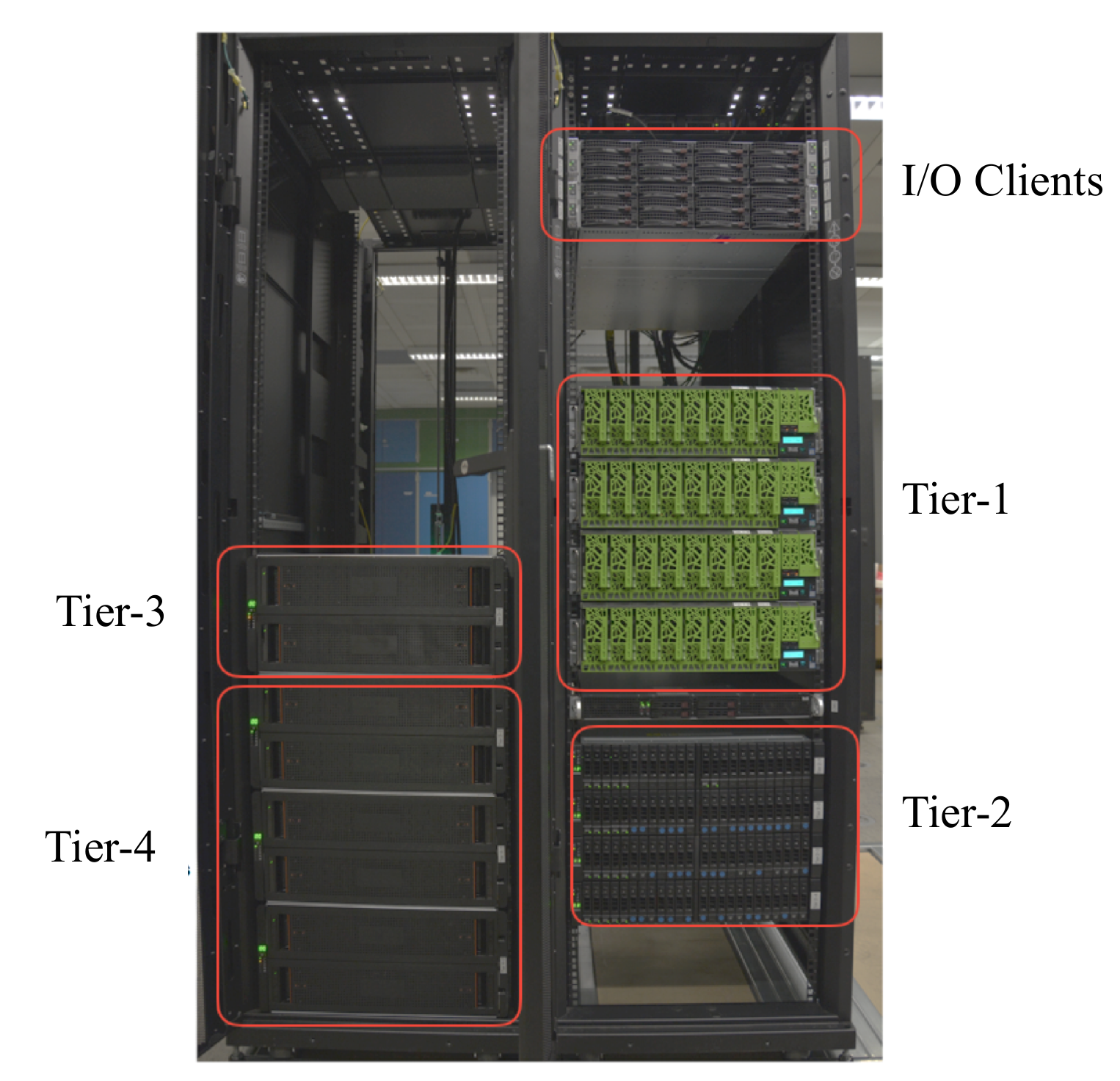}
  \caption{SAGE supercomputer prototype that has been installed at the J\"ulich Supercomputing Center in 2017.}
 \label{fig:Figure1}
 \end{center} 
\end{figure}

A SAGE prototype system has been developed by Seagate and other SAGE consortium partners and is installed at the at the J\"ulich Supercomputing Center. Figure \ref{fig:Figure1} shows the SAGE prototype consisting of the four tiers:
\begin{itemize}
\item Tier-1: PCIe-attached NVMe SSDs based on NAND Flash or 3D XPoint memory
\item Tier-2: SAS-attached SSDs based on NAND Flash
\item Tier-3: High performance disks
\item Tier-4: Archival grade disks.
\end{itemize}

\section{SAGE Software Stack}
\label{sec-architecture} 
\begin{figure}
  \begin{center} 
  \includegraphics[width=0.9\linewidth]{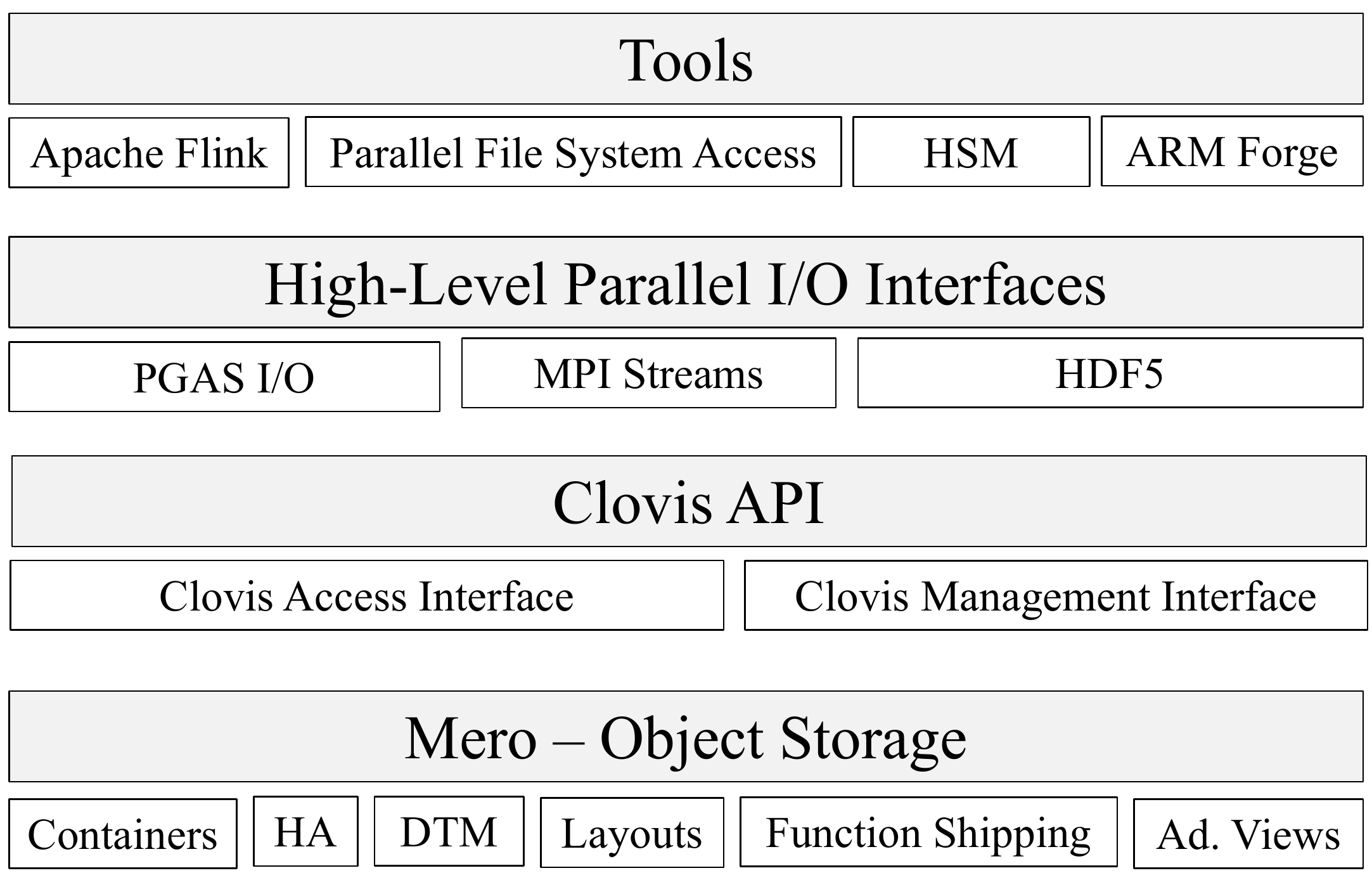}
  \caption{SAGE Software Stack.}
 \label{fig:Figure2}
 \end{center} 
\end{figure}

Together with the development of SAGE storage centric prototype platform, we designed and developed software capable of taking advantage of the SAGE infrastructure. A simplified diagram of the SAGE software stack is presented in Figure \ref{fig:Figure2}.

\subsection{Mero}
Mero is at the base of the SAGE software stack, providing the Exascale-capable object storage infrastructure that drives the storage hardware \cite{danilov2016mero}. Mero is a distributed object storage platform. Lustre file system, NFSv4 and database technology are the the main sources of inspiration for the Mero design. An emerging consensus is that traditional file system properties (hierarchical directory namespace, strong POSIX consistency guarantees, etc) pose a major obstacle for the performance and scalability of Exascale systems. 

Mero controls a cluster of nodes connected by a network. Some nodes have persistent storage attached to them. Mero distinguishes various types of persistent stores including rotational drives, SAS-attached non-volatile memory devices and PCIe-/memory bus attached non-volatile memory devices. Mero presents a collection of services to applications.

Mero Object store has a "core" providing - scalable re-writable fault-tolerant data objects, Index store with scalable key-value indices, and, resource management capabilities for caches, locks, extents, etc. 

\emph{Extreme scale} features such as containers and function shipping are built on top of the core. Mero also features a Distributed transaction manager, which makes it possible to use other services in a consistent manner in the face of hardware and software failures, which is also an extreme scale feature on top of the core. 

Mero also has an Extension mechanism which allows the addition of new functionalities without modification to the core, called the FDMI, or File Data Manipulation Interface. New services such as Information lifecycle management~(ILM), indexing, search, etc can hence be built as extensions by third parties leveraging the core. 

A Mero application can be a traditional user space application, running standalone on a node, a large MPI job running on multiple nodes, a cluster management utility monitoring the state of the system, or a NFS/CIFS daemon (a front-end) exporting Mero objects to non-Mero clients, etc.

\textbf{Container abstraction.} Containers are the main way of grouping objects in various ways. They virtualize the object name space by providing the capability to label objects in various ways. There can be containers based on data formats (for example, HDF5 containers, etc) and containers based just on performance (for example, high performance containers which are mapped to objects in higher tiers, etc). It is possible to do operations such as function shipping, pre/post processing on a given containers.

\textbf{High Availability (HA) System.}  Existing systems and data ~\cite{failures} indicate that we can expect many hardware failures per second at Exascale, in addition to software failures resulting in crashed nodes. To maintain service availability in the face of expected failures, a global state (or configuration) of the cluster is maintained. This may need to be modified by the means of repair procedures in response to failure events. The HA subsystem for SAGE will perform such automated repair activities within storage device tiers. The HA subsystem thus monitors failure events (inputs) throughout the storage tiers and then decides to take action based on collected events. 

\textbf{Distributed Transaction Management (DTM).} In Mero, all I/O and metadata operations are, ultimately, organized into transactions. Transactions are atomic with respect to failures. In other words, either all or none of the updates corresponding to a transaction are visible to other users. This property is known as atomicity. A related property is failure atomicity, i.e. either all or none of the updates corresponding to a transaction survive a failure. A group of updates that must atomically survive a failure is called a distributed transaction.

Mero implements a Distributed Transaction Manager (DTM) that guarantees efficient management of system state consistency in an environment in which dependent data and metadata are scattered over multiple nodes to provide fault tolerance and scalability. DTM provides the necessary interface to group a collection of object store updates into a distributed transaction and guarantees that, in the event of a server node failure and restart, the effects of distributed transactions that have updates for the affected server are either completely restored after restart or completely eliminated.

\textbf{Layouts.} 
A layout is a mapping of different parts or regions of an object to storage tiers. Each object has a layout attribute that defines how the object is stored in the cluster. Mero provides a flexible, hierarchical description to map object subcomponents to physical locations, or storage tiers. This mapping allows for compact formulaic expressions,  as well as data transformations, such as erasure coding, de-duplication, encryption and compression. Layouts also describe data redundancy models, like simple replication or Server Network Striping. As an example of a layout, an object can have some extents mapped to Tier-1, other extents mapped to Tier-2 and a few others mapped to Tier-3. Further, each of set of extents mapped to a certain tier can have its own "sub-layout".

\textbf{Function Shipping.} Function shipping in Mero provides the ability to run application functions directly on storage nodes. This addresses one of the big bottlenecks foreseen for Exascale systems, which is the overhead of moving data to computations. Indeed moving very large quantities of data from storage to compute is extremely energy intensive and energy is one of the prime candidates to address for Exascale systems. Well defined functions within the use cases are registered on the storage nodes and are invoked by the use cases using remote procedure calls. Function shipping is accomplished through extensions of the Mero Clovis API (see next section).

\textbf{Advanced Views.} Modern storage systems are increasingly heterogeneous. This means not only multitude of data sources and data formats, but also the multitude of applications accessing shared data sets with different access patterns, multitude of various storage policies (retention, access control, provenance, etc.) and multitude of conflicting goals that the storage system must balance (latency vs. throughput, different storage hardware).
Historically, providing an interface that allowed different applications to access shared data often resulted in great benefits both for application and system developers, the two most famous examples being UNIX "everything-is-a-file" and relational models. Data stored in Mero are accessible to multiple applications using various existing storage interfaces (e.g., POSIX, pNFS, S3, HDF5). A component of Mero called \emph{Lingua Franca} (LF) implements common meta-data formats and interfaces that enables interoperability between multiple external interfaces and internal meta-data users.

LF is a mechanism to share the same sets of storage entities (objects, indices and containers) between multiple applications with different access interfaces. Its straightforward use is to provide interoperability between different front-ends. Together with other Mero features, like containers and function shipping, it can be used to implement a very flexible access arrangement to the data.



\subsection{Clovis}

Mero's application programming interface (API), known as Clovis, provides a library of functions that applications and front-end programs can use for accessing and manipulating storage resources in Mero. Access to storage resources by outside applications is strictly controlled via Clovis; no other interfaces exist. The Clovis API contains optimized functions to manage performance and scalability for modern extreme scale computing applications as well as legacy applications. We expect higher-level applications, as part of development work related to accessing Mero storage resources, to build their own APIs on top of Clovis. In other words, we do not expect computational scientists to directly interface with the Clovis interface. Higher-level interfaces include HDF5, POSIX via pNFS, and others.

The Clovis API is implemented in the C programming language, but equivalent versions of the API are planned in other popular programming languages such as Java, Python, etc. The Clovis API supports asynchronous transactions, i.e. an application can continue after starting a transaction and only later check for completion of the transaction. For clarity, the API is divided into three sub-APIs: the Clovis Access API, the Clovis Management API and the Clovis Extension API. 

\begin{figure}
  \begin{center} 
  \includegraphics[width=0.9\linewidth]{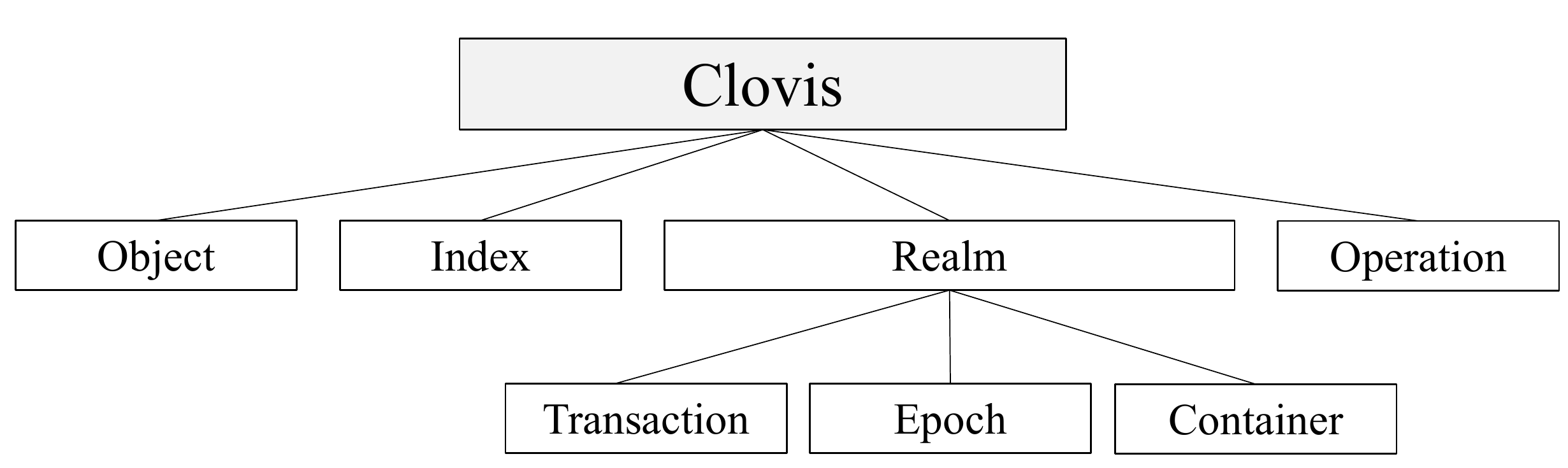}
  \caption{Clovis abstractions.}
 \label{fig:Figure1}
 \end{center} 
\end{figure}

Clovis provides the following abstractions, shown in Figure~\ref{fig:Figure1}. Short descriptions of the abstractions are provided below:	
\begin{itemize}					 							
\item Object is an array of fixed size blocks of data
\item Index is a key-value store 
\item An entity is an object or an index							
\item Realm is a spatial and temporal part of the system with a prescribed access discipline
 \item Operation is a process of querying and/or updating the system state
\item Objects, indices and operations live in realms	
\item Transaction is a collection of operations that are atomic in the face of failure
\item Epoch is a collection of operations done by an application that moves the system from one application-consistent state to another
\item Container, in the Mero context, is a collection of objects used by a particular application or group of applications 					
\end{itemize}	
					
The Clovis Access API handles all I/O related functions for applications. In addition to standard initialization and finalization functions related to running a Clovis instance, the API provides \textsf{create()}, \textsf{write()}, \textsf{read()}, and \textsf{free()} functions that enable applications to create Mero objects, transfer data to and from them, and then delete them (completely freeing the resources the objects use). 

The Clovis Management API handles all management-related functions in Mero, including system configuration, management of services, and analytics as well as diagnostics functions.	
								
The Clovis Extension API provides a list of compute functions that support the development of third-party Mero plug-ins without modifying the core. The FDMI (File Data Manipulation Interface) is an example for how to use this feature.

\subsection{Higher-Level I/O Interfaces}
At the top of the software stack, we further develop widely-used HPC legacy APIs, such as MPI and HDF5, to exploit the SAGE architecture.

\textbf{PGAS I/O.} The goal of the Partitioned Global Address Space (PGAS) programming model is to provide processes with a global view of the memory and storage space during the execution of a parallel application. This is similar to what a Shared Memory model provides in a multithreaded local environment. In the PGAS approach, remote processes from different nodes can easily collaborate and access memory addresses through load / store operations that do not necessarily belong to their own physical memory space. In SAGE, we propose an extension to the MPI one-sided communication model to support window allocations in storage: MPI storage windows~\cite{rivas2017extending,rivas2017mpi}. Our objective is to define a seamless extension to MPI to support current and future storage technologies without changing the MPI standard, allowing to target either files (i.e., for local and remote storage through a parallel file system) or alternatively address block devices directly (i.e., as in DRAM). We propose a novel use of MPI windows, a part of the MPI process memory that is exposed to other MPI remote processes, to simplify the programming interface and to support high-performance parallel I/O without requiring the use of MPI I/O. Files on storage devices appear to users as MPI windows (MPI storage windows) and seamlessly accessed through familiar \textsf{PUT} and \textsf{GET} operations. Details about the semantics of operations on MPI storage windows and the implementation are provided in Ref.~\cite{rivas2017mpi}.

\textbf{MPI Streams for Post-Processing and Parallel I/O.} While PGAS I/O library addresses the challenge of heterogenous storage and memory, streams can be used to support function-shipping for post-processing and highly scalable parallel I/O. {\em Streams} are a continuous sequence of fine-grained data structures that move from a set of processes, called data {\em producers}, to another set of processes, called data {\em consumers}. These fine-grained data structures are often small in size and in a uniform format, called a {\em stream element}. A set of computations, such as post-processing and I/O operations, can be {\em attached} to a data stream. Stream elements in a stream are processed {\em online} such that they are discarded as soon as they are {\em consumed} by the attached computation. 

In particular, our work in SAGE focuses on {\em parallel streams}, where data producers and consumers are distributed among processes that require communication to move data. To achieve this, we have developed a stream library, called MPIStream, to support post-processing and parallel I/O operations on MPI consumer processes~\cite{peng2017mpi, peng2017preparing,markidis2016performance}. More details about MPI streams operation semantics and MPIStream implementation are provided in Ref.~\cite{peng2015data}. 

\textbf{HDF5.} Typically, data formats in HPC provide their own libraries to describe data structures and their relations (including I/O semantics). The HDF5 data format  needs to be supported in SAGE, and is layered directly on top of Clovis. The HDF5 will use the Virtual Object Layer Infrastructure within HDF5 (used to interface HDF5 with various object formats), to interface with Clovis.

\subsection{Tools}
A following are a set of tools for I/O profiling and optimized data movement across different SAGE platform tiers at the top of the SAGE software stack. 

\textbf{Data Analytics Tools.} Apache Flink is a framework for data analytics workloads. Flink connectors for Clovis are currently under development to enable the deployment of data analytics jobs on top of Mero. 

\textbf{Parallel File System Access.} Parallel file system access is the traditional method of accessing storage in HPC. Many of the SAGE applications and use cases  need the support of POSIX compliant storage access. This access is provided through the pNFS gateway built on top of Clovis. However, pNFS requires some POSIX semantics to be developed by leveraging Mero's KVS. For instance, an abstraction of namespaces on top of Mero objects is needed.

\textbf{Hierarchical storage management and Data Integrity.} In SAGE, an Hierarchical Storage Management (HSM) is used to control the movement of data in the SAGE hierarchies based on data usage. Advanced integrity checking overcomes some of the drawbacks of well known and widely used file system consistency checking schemes. 

\textbf{ARM Forge.} ADDB telemetry records from the Clovis management interface are directly fed to ARM Forge performance report tools that reports overall system performance for SAGE.

\section{Validation with Applications}
As seen in the previous section, the SAGE platform supports appropriate scientific computing data formats and legacy application interfaces such as parallel file systems and POSIX. SAGE also needs to interface with emerging big data analytics applications (on top of the API) to access the rich features of these tools, and the Volumes, Velocity and Variety (potentially) of data coming from sensors, instruments and simulations. We have created a portfolio of scientific data-centric applications that have been used to provide requirements to the development of the SAGE system and to validate the developments in the projects. The applications we chose for the SAGE project are:
\begin{itemize}
\item  iPIC3D is a parallel Particle-in-Cell Code for space physics simulations in support of NASA and ESA missions \cite{markidis2010multi, peng2015energetic, peng2015formation}.
\item  NEST is a spiking neural network simulator to study brain science \cite{gewaltig2007nest}.
\item  Ray is a parallel meta-genome assembler \cite{boisvert2012ray}.
\item JURASSIC is a fast radiative transfer model simulation code for the mid-infrared spectral region \cite{griessbach2013scattering}.
\item EFIT++ is a plasma equilibrium fitting code with application to nuclear fusion \cite{lupelli2015efit++}.
\item The ALF code performs analytics on data consumption log files
\item Spectre is a visualization tool providing near real time feedback on plasma and other operational conditions in fusion devices.
\item Savu is a code for tomography reconstruction and processing pipeline \cite{wadeson2016savu}.
\end{itemize}

\section{Related Work}
\label{sec-relwork}
To the best of our knowledge SAGE is the first HPC-enabled storage system to implement new NVRAM tiers, flash and disk drive tiers as part of a single unified storage system. The SAGE Architecture progresses the state of the art from Blue Gene Active Storage~\cite{fitch2010blue} and Dash~\cite{he2010dash}, which use flash for data staging. It also progresses the state of the art from Burst Buffer technologies as discussed earlier. 

When compared to the FastForward Project~\cite{FastForward}, SAGE highly simplifies storage, developing a solution for deep I/O hierarchies, including NVRAM technologies. A major difference between SAGE and FastForward is the FastForward solution is evolutionary as it tries to make use of an existing storage solution, namely, Lustre~\cite{schwan2003lustre} used for the last 20 years or so. Lustre was really designed for the previous era, when use cases and architectural assumptions were different. On the hand, SAGE and Mero are the product of a complete redesign in consideration of the new requirements arising out of the Extreme scale computing community. 

Mero, the object store in SAGE, extends the state of the art in existing object storage software such as Ceph~\cite{weil2006ceph} and Open Stack Swift ~\cite{swift} by building Exascale components required for extreme scale computing. While Ceph and Openstack swift are designed for supporting mainly cloud storage, Mero is built to meet the needs of the extreme scale computing community.

\section{Conclusions}
\label{sec-conclusions}
The SAGE project objective is to design and implement an I/O system capable of supporting I/O workloads of Exascale supercomputers. The SAGE platform has been recently installed at J\"ulich Computing center. It supports a multi-tiered I/O hierarchy and associated software stack, to provide a demonstrable path towards Big Data analytics and HPC convergence. The SAGE software stack consists of four main software layers: the Seagate Mero object-storage, the Clovis API, high-level interfaces and tools. 

Current ongoing work focuses on the performance characterization of various new NVRAM device technologies. We also currently investigating lower level software and Operating System~(OS) infrastructure requirements to exploit these new devices types, below Mero in the SAGE stack. We clearly recognize that various NVRAM technologies have their own performance characteristics and limitations. New NVRAM technologies can be part of the SAGE hardware tiers based on where they ultimately are on the performance and capacity curve. The SAGE stack and Mero indeed is designed to be agnostic of storage device types as long as adaptations are in place within the  OS. 

The next steps will be to quantify the benefits of the various features of the SAGE stack on the SAGE prototype system currently installed at J\"ulich  Supercomputing Center, with focus on providing results for the remaining SAGE components and the SAGE architecture as a whole. As a part of this external organizations outside of the SAGE consortium ( eg: from Climate and Weather, Astronomy, etc) will soon be granted access to study how their codes and workflows can exploit the features of the SAGE platform.  We will then look at extrapolation studies of the benefits of the various SAGE features at Exascale through analytical and simulation models. These will be discussed separately. Porting of the SAGE stack across other sites and extensions of the SAGE prototype is also planned. We are targeting SAGE work to be a part of European Extreme Scale Demonstrators~\cite{esd} which will be pre-Exascale prototypes. 

\section*{Acknowledgements}
The authors acknowledge that  the SAGE work is being performed by a consortium of members consisting of Seagate (UK), Bull ATOS (France), ARM (UK), KTH (Sweden), STFC (UK), CCFE (UK), Diamond (UK), DFKI (Germany), Forchungszentrum J\"ulich (Germany) and CEA (France) - which is being represented by the Authors in this paper. 

Funding for the work is received  from the European Commission H2020 program, Grant Agreement No. 671500 (SAGE). 
\bibliographystyle{ACM-Reference-Format}
\bibliography{SAGE-storage-centric}

\end{document}